\pdfoutput=1

\documentclass[12pt]{iopart}

\usepackage{newtxtext}

\usepackage[T1]{fontenc}
\usepackage{graphicx}
\usepackage{booktabs}
\usepackage{multicol}
\usepackage{subcaption}
\usepackage{array}
\usepackage{geometry}

\expandafter\let\csname equation*\endcsname\relax
\expandafter\let\csname endequation*\endcsname\relax
\usepackage{amsmath}
\usepackage{natbib}

\newcommand{\gray}{$\gamma$-ray\ }
\newcommand{\grays}{$\gamma$-rays\ }

\begin{document}

\title[Layout optimization and performance of LACT]{Layout optimization and Performance  of Large Array of  imaging atmospheric Cherenkov Telescope (LACT)}
\author{
Zhipeng Zhang$^{1}$,
Ruizhi Yang$^{\star 1,2,3,6}$,
Shoushan Zhang$^{4,5,6}$,
Zhen Xie$^{1}$,
Jiali Liu$^{4,5,6}$,
Liqiao Yin$^{4,5,6}$,
Yudong Wang$^{4,5,6}$,
Lingling Ma$^{4,5,6}$,
and Zhen Cao$^{4,5,6}$,
}

\address{$^{1}$School of Astronomy and Space Science, University of Science and Technology of China, Hefei, Anhui 230026, China}
\address{$^{2}$CAS Key Laboratory for Research in Galaxies and Cosmology, Department of Astronomy, University of Science and Technology of China, Hefei, Anhui 230026, China}
\address{$^{3}$Deep Space Exploration Laboratory/School of Physical Sciences, University of Science and Technology of China, Hefei 230026, China}
\address{$^{4}$Key Laboratory of Particle Astrophysics, Institute of High Energy Physics, Beijing, China}
\address{$^{5}$Department of Physics, University of Chinese Academy of Sciences, Beijing, China}
\address{$^{6}$Tianfu Cosmic Ray Research Center, Chengdu, China}

\ead{yangrz@ustc.edu.cn}
\vspace{10pt}
\begin{indented}
\item[]June 2024
\end{indented}

\begin{abstract}
Large Array of  imaging atmospheric Cherenkov Telescope (LACT) is an array of 32 Cherenkov telescopes with 6-meter diameter mirrors to be constructed at the LHAASO site. In this work, we present a study on the layout optimization and performance analysis of LACT.  We investigate two observation modes: large zenith angle observations for ultra-high energy events and small zenith angle observations for lower energy thresholds. For large zenith angles (60°), simulations show that an 8-telescope subarray can achieve an effective area of $3 ~\rm km^2$ and excellent angular resolution. For small zenith angles, we optimize the layout of 4-telescope cells and the full 32-telescope array. The threshold of the full array is about $200~\rm GeV$, which is particularly crucial for studying transient phenomena, including gamma-ray bursts (GRBs) and active galactic nuclei (AGNs). This study provides important guidance for the final LACT layout design and performance estimates under different observational conditions, demonstrating LACT's potential for deep observations of ultra-high energy \gray sources and morphological studies of PeVatrons, as well as time-domain \gray astronomy.
\end{abstract}

%
%
%
%
%
\noindent{\it Keywords}: IACT, \gray, LHAASO



\section{Introduction}
{The field of ground-based \gray astronomy has evolved significantly since its inception in the mid-20th century. Since the atmosphere is opaque to \grays, ground-based \gray astronomy is based on the detections of secondaries produced in the interaction between \grays and the atmosphere. In this regard, two primary types of detectors have driven this evolution: Imaging Atmospheric Cherenkov Telescopes (IACTs) and Extensive Air Shower (EAS) arrays. IACTs, such as those developed at the Whipple Observatory \citep{krennrich1998stereoscopic} in the 1960s, detect \grays by capturing the Cherenkov light produced by charged secondary particles resulting from \gray interactions in the Earth's atmosphere. This method allows for high-resolution imaging of \gray sources. On the other hand, EAS arrays, like the Tibet AS-$\mathrm \gamma$ \citep{amenomori1992search} and ARGO-YBJ \citep{bacci2002results}, directly detect the secondary particles that reach the ground. These arrays offer a broad field of view and are particularly effective for surveying extensive portions of the sky. }

{Over the past 20 years, IACTs have fundamentally led the field of TeV \gray astronomy. The current generation IACTs, such as H.E.S.S \citep{aharonian2006observations}, Veritas \citep{veritas2011detection}, MAGIC \citep{aleksic2012performance} have made numerous significant contributions, greatly enhancing our understanding of high-energy phenomena in the universe. More than three hundred TeV sources have been detected so far, most of which are attributed to the IACTs \footnote{http://tevcat.uchicago.edu/}. }

{Nowadays, the landscape of \gray astronomy is being increasingly shaped by  LHAASO (Large High Altitude Air Shower Observatory \citep{cao2019large}). 
 As a leading facility in EAS arrays, LHAASO has pioneered the field of ultra-high energy (UHE) \gray astronomy with its unprecedented sensitivity above $20 ~\rm TeV$. It opened the window of UHE \gray astronomy by detecting the first 12 UHE \gray sources in the Galactic plane \citep{cao2021ultrahigh}. }



Recently, LHAASO has published its first catalog (1LHAASO \citep{cao2024first}), which includes over 90 high-energy \gray sources. Notably, 43 of these sources have energies exceeding $100 ~\rm TeV$. These sources are candidates of PeV particle accelerators, which are dubbed PeVatrons and are crucial for understanding the origin of Cosmic rays (CRs) in our Galaxy. However, most of these sources are extended, and due to LHAASO's limited angular resolution, accurately identifying the origins of these ultra-high energy \gray emissions remains challenging \citep{cao2023ultra}. Compared to LHAASO, IACT arrays can provide much better angular resolution, but the effective area of current IACT arrays is only around  $ 10^5~\rm 
 m^2$, which makes it hard to have good synergy with LHAASO. 
The next generation of Cherenkov telescope arrays, such as CTA (Cherenkov Telescope Array \citep{acharya2013introducing}) and ASTRI \citep{vercellone2022astri}, will have effective areas greater than $ 10^6 ~\rm m^2$. With their excellent angular resolution and larger effective area, CTA and ASTRI will provide a powerful complement to LHAASO. In addition, we propose LACT (Large Array of imaging atmospheric Cherenkov Telescope \citep{Zhang:2024+q}) at the LHAASO site as a crucial advancement in this direction, offering significantly improved sensitivity and angular resolution for detailed studies of LHAASO-detected sources, particularly in the ultra-high energy range. 


LACT consists of 32 telescopes, each with a $6~\rm m$ diameter.  These telescopes will employ SiPM technology for their cameras, a technology already well-validated on WFCTA \citep{aharonian2021construction}. This advancement will enable telescopes to operate during moon nights, significantly increasing the observation time. The primary scientific goal of LACT is to conduct long-term observations of PeVatrons discovered by LHAASO, utilizing its superior angular resolution to study the morphology of these sources. Additionally, LACT aims to perform well at energies below $1 ~\rm TeV$, enabling it to observe extragalactic sources and \gray transients discovered by LHAASO-WCDA \citep{hu2023first}. This can significantly broaden the scientific objectives of LACT. Based on these considerations, two different observation modes have been proposed: one for large zenith angle observations targeting ultra-high energy events, and the other for normal small zenith angle observations. The optimization process for LACT's layout must take both of these factors into account. Given the complexity of telescope layouts and the need to explore baseline performance in both observation modes, we conducted this study to guide the final layouts and estimate the performance of LACT under different observational conditions. 

{This paper is organized as follows: In Section 2, we briefly introduce the Monte Carlo simulation and reconstruction methods. In Section 3, we investigate the layout and performance in large zenith angle observations. In Section 4, we discuss the small zenith angle situation, beginning with an individual cell and then examining the performance of the entire array. Finally, in Section 5, we conclude our results and discuss the implications.}

\section{SIMULATION AND RECONSTRUCTION METHODS}

We used the CORSIKA \citep{heck1998corsika} package (version 7.64) to produce the \grays and proton air showers. For electromagnetic interactions, we employed the EGS4 model, while for hadronic interactions, we used the URQMD model at low energies and the QGSJET-II model at high energies. The photon files obtained from CORSIKA were piped into sim\_telarray \citep{bernlohr2008simulation} to generate the telescope's response. {The telescope configuration used in the simulation consists of a Davies-Cotton design mirror with a $6~\rm m$ diameter and an $8 ~\rm m$ focal length. The telescope's camera is composed of over 1,400 pixels, each with a size of 
$25.8~\rm mm$, resulting in a total field of view of $8^{\circ}$ in diameter. In this simulation, we generated events for point-like gamma rays, diffuse gamma rays, and diffuse protons at zenith angles of 20° and 60°. The diffuse gamma rays and diffuse protons were randomly distributed within a cone with a 7° radius centered on the position of the simulated point-like source. To increase the number of showers, we reused shower events: point-like gamma events were reused 10 times, while diffuse gamma and diffuse proton events were reused 20 times each.
The specific parameters of the simulation can be found in Table 1.} 

For each event, we required at least two telescopes to trigger. We first performed image cleaning on the telescope images: a two-level tail-cut image cleaning method was employed, which requires a pixel to be above a specified high threshold and at least one of its neighboring pixels to be above a low threshold, or vice versa \citep{bernlohr2013monte}. After image cleaning, the image was parameterized \citep{hillas1985cerenkov}. 
In addition to the standard Hillas parameters, we also introduced the \textit{MISS} parameter, defined as the distance from the true source position to the major axis in the nominal plane. The \textit{MISS} parameter can represent the reconstruction accuracy of the shower-detector plane (SDP) for a single telescope, and we will frequently refer to it in the following sections. For reconstruction, we required at least two telescopes to pass the following selection cut:  \textit{SIZE} > 100 photoelectrons(p.e.), and \textit{LEAKAGE2} < 0.3, where \textit{SIZE} is the total p.e. in the image after cleaning, and  \textit{LEAKAGE2} is the ratio of p.e. in the outermost two layers of pixels. {It is worth noting that these selection cut conditions have not been optimized and are only preliminary.}

The direction of the incoming shower was reconstructed by the intersection of major axes in the reference telescope frame. After reconstructing the direction and core position of each event, we calculated the corresponding reconstructed Impact Parameter. Combining this with the parameters obtained from the telescopes, we trained a RandomForestRegressor model for energy reconstruction and a RandomForestClassifier model for particle separation \citep{scikit-learn} using diffuse gamma events and diffuse proton events. 
The estimated energy and hadroness of a single telescope were combined with weights to determine the overall reconstructed energy and hadroness of the event. To facilitate easier comparisons, we will typically compare the angular resolution and collection area after event selection versus the true energy in the following sections.

\section{LARGE ZENITH ANGLE OBSERVATION}
Increasing the effective area of Cherenkov telescopes at high energies by observing at large zenith angles (LZA) was proposed early on \citep{konopelko1999effectiveness} and
it has been widely applied to existing  IACTs:
MAGIC, by using Very Large Zenith Angle observation(zenith angle $70^{\circ}\sim80^{\circ}$) mode, increased the collection area to 2 $\rm km^2$ and successfully detected the Crab Nebula's spectrum up to $100 ~\rm TeV$ \citep{acciari2020magic};
{In addition to increasing the effective area, observations at large zenith angles can also increase  the sky coverage of IACTs.
 VERITAS has  studied the Galactic Center region, which can only be observed with LZA in VERITAS site} \citep{adams2021veritas}. Similary, for LACT, during the suitable observation period, the Galactic Center (RA: 17h 45m 39.6s, DEC: -29° 00' 22") can only be observed at zenith angles above 50°.
 
\begin{figure}
    \centering
    \begin{subfigure}[b]{0.5\textwidth}
        \centering
        \includegraphics[width=\textwidth]{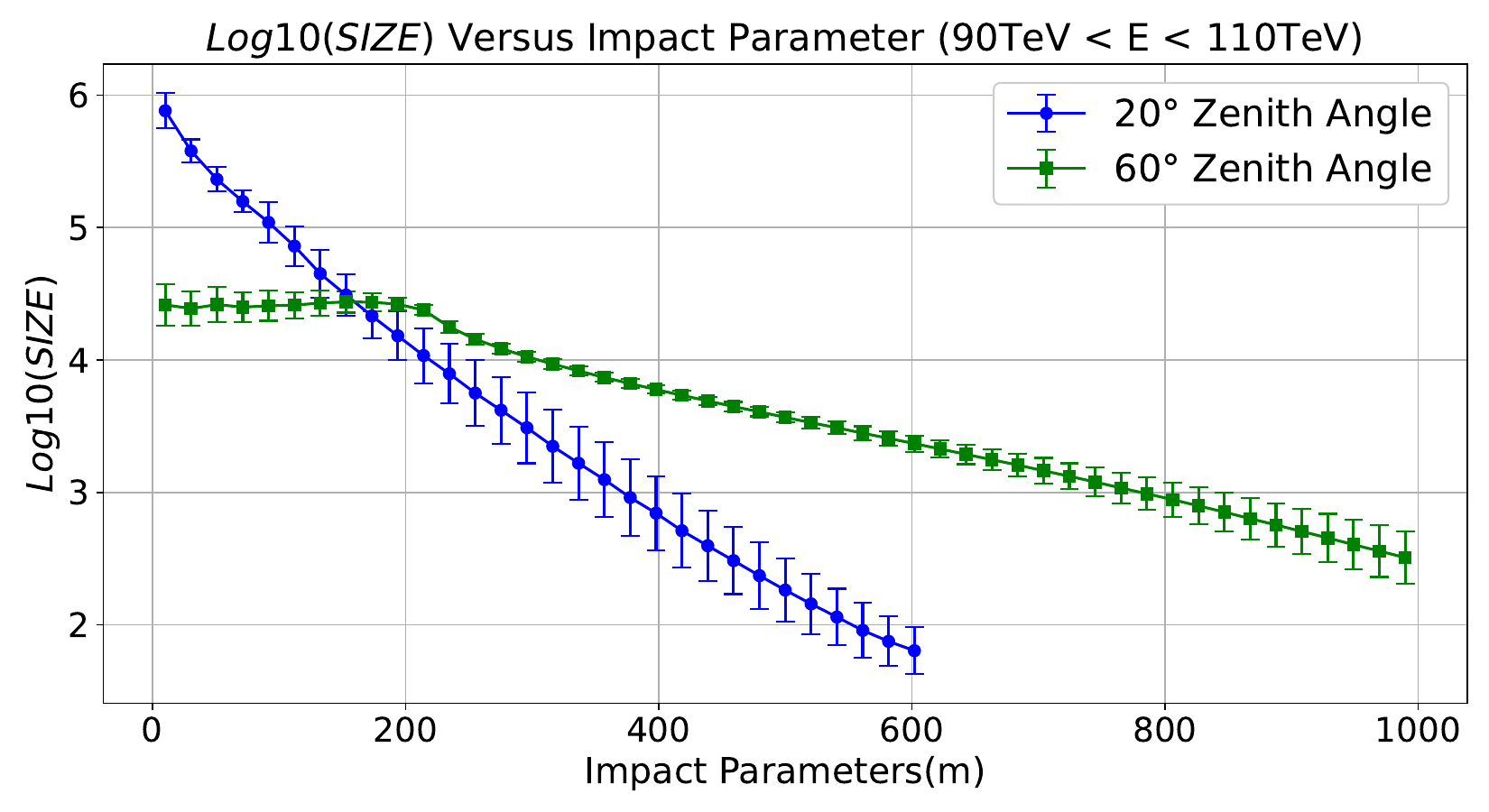}
        \caption{The total number of photoelectrons in the telescope after image clean versus the impact parameter.}
        \label{fig1:a}
    \end{subfigure}
    \hfill
    \begin{subfigure}[b]{0.5\textwidth}
        \centering
        \includegraphics[width=\textwidth]{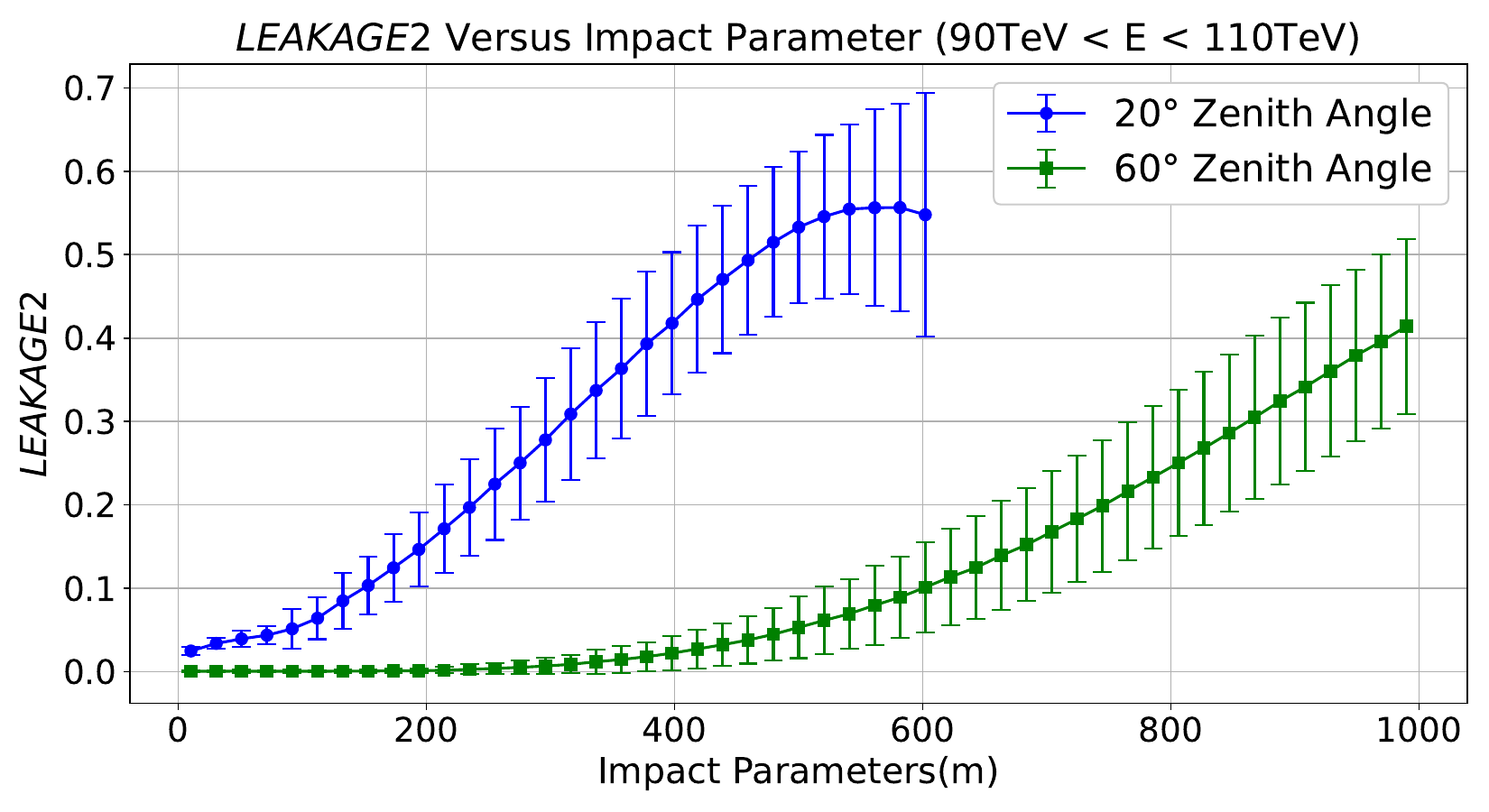}
        \caption{\textit{LEAKAGE2} versus impact parameter.}
        \label{fig1:b}
    \end{subfigure}
    \hfill
    \begin{subfigure}[b]{0.5\textwidth}
        \centering
        \includegraphics[width=\textwidth]{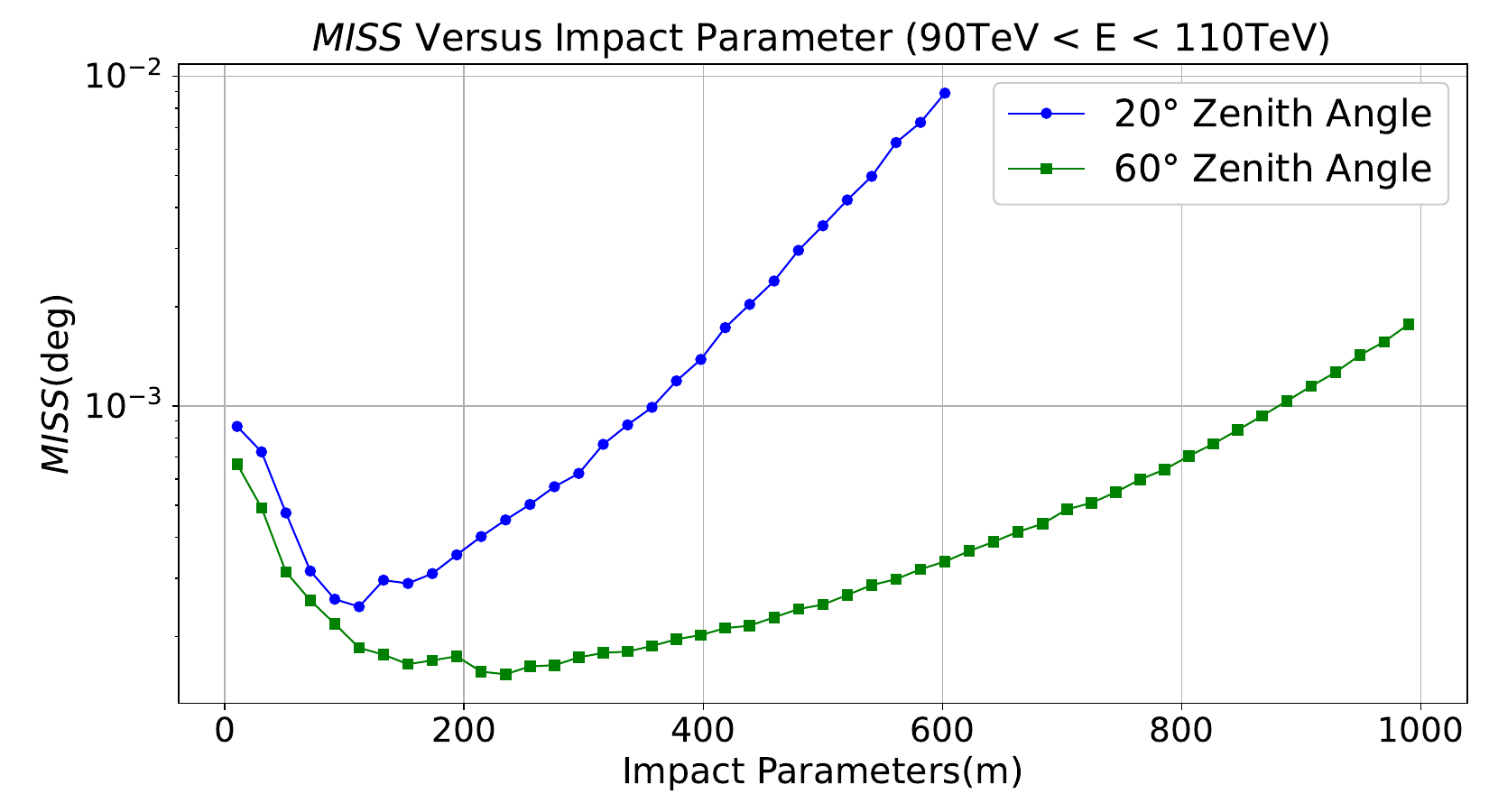}
        \caption{ Mean \textit{MISS} versus impact parameter.}
        \label{fig1:c}
    \end{subfigure}
    \caption{Comparation of some parameters for small zenith angle ($20^{\circ}$) and large zenith angle ($60^{\circ}$).}
    \label{fig:both}
\end{figure}

 Compared to the existing IACTs, which are typically located at altitudes around 2000 meters, the LHAASO site, at a higher altitude of 4400 meters, benefits even more from the enhancements provided by large zenith angle observations. At the LHAASO site, the shower maximum for $100 ~\rm TeV$ gamma-ray showers is very close to the ground, resulting in a very steep photon lateral distribution, which can lead to significant image leakage in the telescopes. However, in the LZA observation mode, the increased atmospheric depth between the shower maximum and the telescope flattens the lateral distribution of photons, resulting in smaller and better-quality images. In Figure \ref{fig:both}, we showed the relationships between the observed \textit{SIZE} and impact parameter, as well as \textit{LEAKAGE2} and impact parameter, for \gray showers around $100 ~\rm TeV$ at $20^{\circ}$ and $60^{\circ}$ zenith angles. Figure \ref{fig1:a} shows that at a $60^{\circ}$ zenith angle, the telescope's image still has a significant number of p.e. even with an impact parameter greater than 800 meters. In contrast, at smaller zenith angles, the steeper lateral distribution limits the detectable distance of the telescope.
Considering the cuts used in our analysis: \textit{SIZE} > 100 p.e. and \textit{LEAKAGE2} < 0.3, the detectable distance of one telescope extends from 300 meters in low zenith mode to over 800 meters in the LZA mode. This significantly increases our collection area and multiplicity.

\begin{figure}
    \centering
    \includegraphics[width=0.5\textwidth]{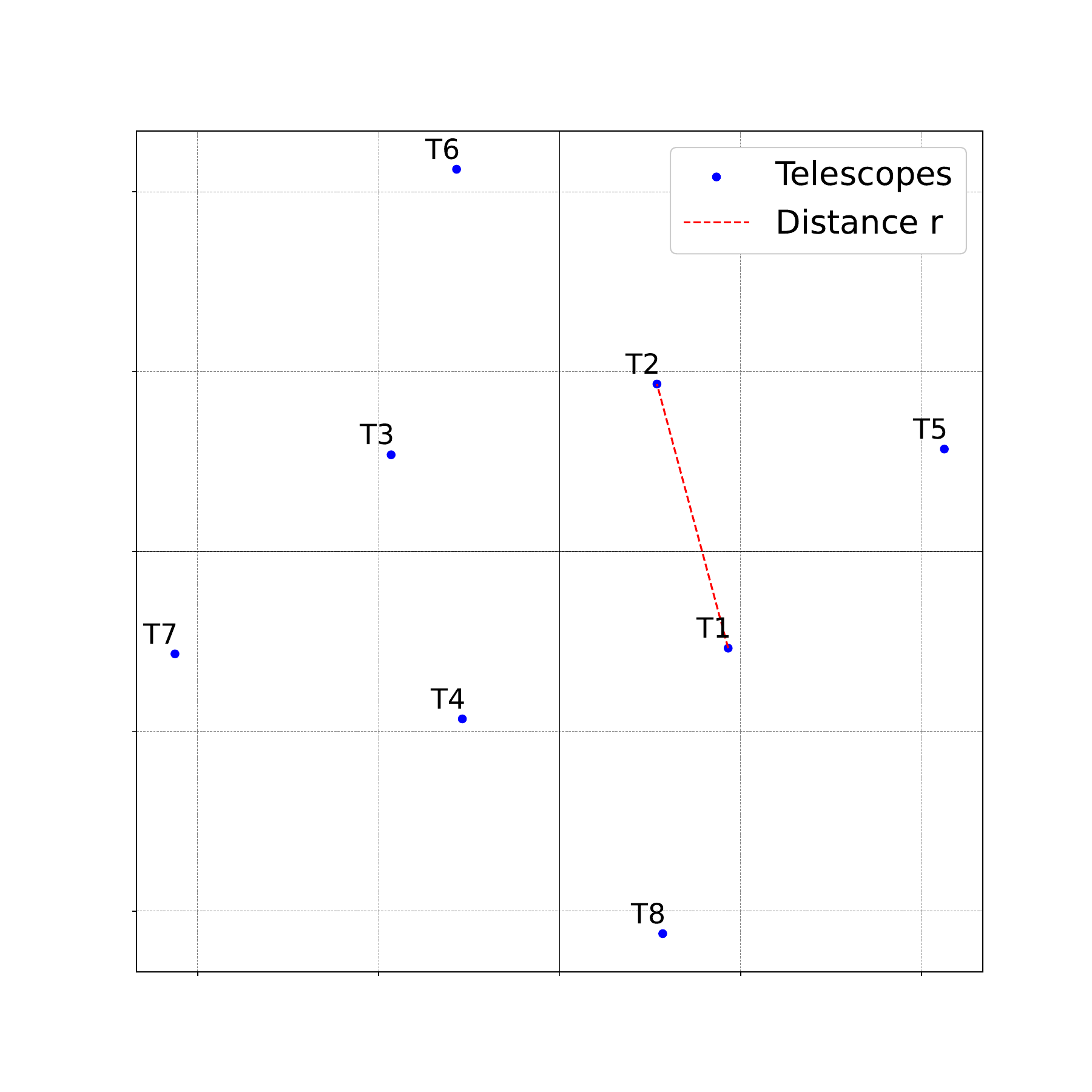}
    \caption{Telescope Layout for Eight Telescopes. The line between Tel.1 and Tel.2 represents the distance \( r \).}
    \label{fig:telescope_layout}
\end{figure}

Although LZA mode can significantly increase the effective area, the angular resolution at LZA for existing IACTs is worse (> $0.1^{\circ}$). This limitation arises because the distance between telescopes in existing IACT arrays is around 100 meters. When observing distant events at large zenith angles, the images captured by different telescopes are nearly parallel, making effective stereoscopic reconstruction challenging \citep{lu2013improving}. Based on the above considerations, we propose to divide the 32 telescopes of LACT into 4 groups for LZA observations, with each group consisting of 8 well-separated telescopes. 
To ensure similar performance for each group, we divided the 32 telescopes into 8 cells, each composed of four closely spaced telescopes. Under LZA observations, the eight telescopes from different cells can be combined to form four groups. This arrangement allows each group to maintain optimal performance in LZA mode. 
To further investigate whether larger distances between telescopes improve performance at large zenith angles, we conducted the following studies. 
Figure \ref{fig:telescope_layout} shows the layout of eight telescopes, where 
r is the distance between each telescope and its nearest neighbor. In the simulation, we considered situations with 
r values of 300, 400, and 500 meters, respectively. The simulated \gray showers have a zenith angle of $60^{\circ}$ and an energy range of $400 ~\rm GeV$ to $400 ~\rm TeV$. 

\begin{table*}
    \centering
    \caption{Simulation parameters}
    \resizebox{\textwidth}{!}{%
    \begin{tabular}{lccccccc}
        \toprule
        Particle type & index & energy range [TeV] & view cone radius [deg] & scatter radius [m] & azimuth direction [deg] & zenith angle [deg] & number of shower \\
        \midrule
        gamma (point-like) & -2 & 0.4--400 & 0 & 1800 & 0 & 60 & $10^8$ \\
        gamma (diffuse) & -2 & 0.4--400 & 7 & 2000 & 0 & 60 & $6\times10^8$ \\
        proton & -2 & 0.6--600 & 7 & 2000 & 0 & 60 & $1.8 \times10^9$ \\
        gamma (point-like) & -2 & 0.1--400 & 0 & 1600 & 0 & 20 & $4 \times 10^8$ \\
        gamma (diffuse) & -2 & 0.1--400 & 7 & 1800 & 0 & 20 & $1.5 \times 10^9$ \\
        proton (diffuse) & -2 & 0.1--600 & 7 & 1800 & 0 & 20 & $4\times 10^9$ \\
        \bottomrule
    \end{tabular}%
    }
\end{table*}
\subsection{The comparison of different layouts}

\begin{figure}
    \centering
    \includegraphics[width=0.5\textwidth]{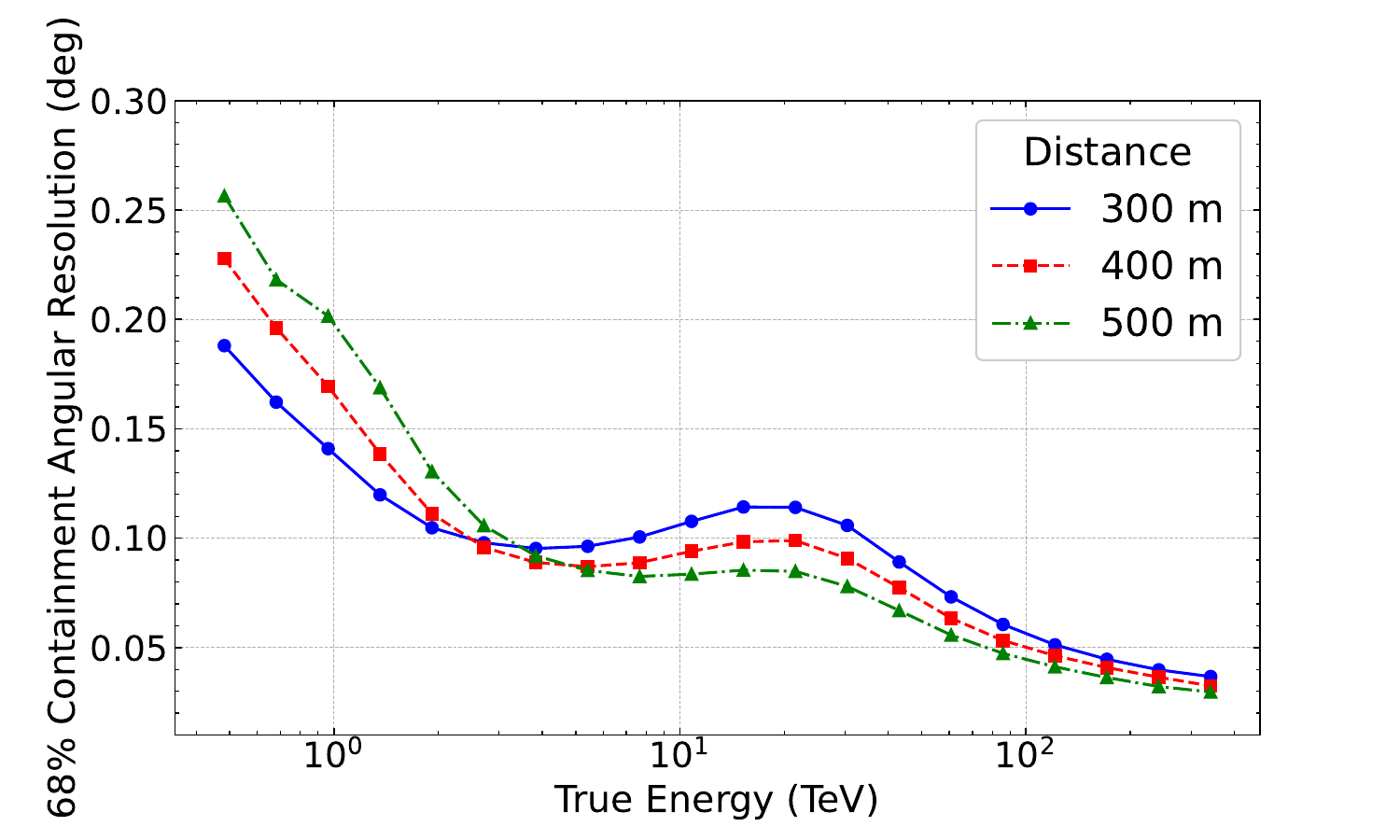}
    \caption{Angular resolution after event preselection at large zenith angle for different distance.}
    \label{fig:largezenith_angres}
\end{figure}

\begin{figure}
    \centering
    \includegraphics[width=0.5\textwidth]{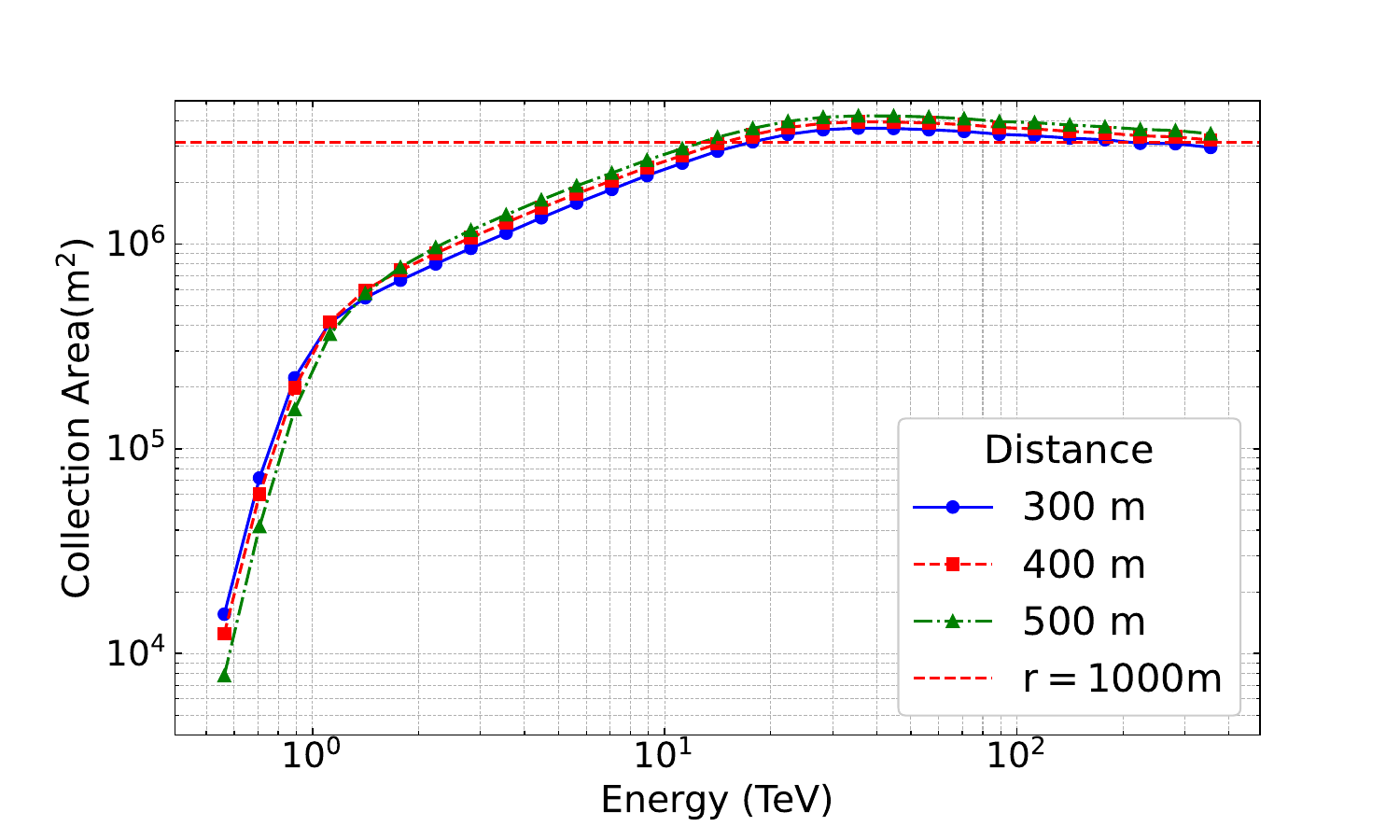}
    \caption{Collection area after event preselection at large zenith angle for different distance. The red dashed line represents the area corresponding to a circle with a radius of 1000 meters.}
    \label{fig:largezenith_collection_area}
\end{figure}

The angular resolution and collection area at different distances are illustrated in Figure \ref{fig:largezenith_angres} and Figure \ref{fig:largezenith_collection_area}. As we expected, above several TeV, the overall performance improves with increasing distance, both in terms of collection area and angular resolution. Considering that the detectable distance of a single telescope (> $800 ~\rm m$) is much greater than the distance between telescopes, extending the distance from $300~\rm m$ to $500 ~\rm m$ can cover a larger area, thereby increasing the collection area. Additionally, owing to that most events are outside the array, increasing the distance between telescopes allows for better stereoscopic reconstruction, thus improving angular resolution. {From the collection area, we observe that it reaches its maximum around $30 ~\rm TeV$. Below $30 ~\rm TeV$, the \textit{LEAKAGE2} cut is less restrictive, resulting in a larger collection area but poorer angular resolution. As the energy increases, the \textit{LEAKAGE2} cut becomes more effective, causing the collection area to gradually decrease with increasing energy.}
Notably, at large zenith angles, the performance at high energies shows significant improvement compared to smaller zenith angles. In former studies \citep{zhang2024prospects}, we have investigated the performance of similar eight telescopes at $20^\circ $ zenith angle. The results show that at large zenith angles, the collection area increases threefold (from $1 ~\rm km^2$ to $3 ~\rm km^2$), accompanied by a substantial improvement in angular resolution, particularly in the energy range above $100 ~\rm TeV$.
{ This enhanced performance at large zenith angles can be attributed to LACT's significantly higher altitude. At lower zenith angles, the air shower's image in the camera is much larger, resulting in significant image leakage even with a larger field of view of $8^{\circ}$. This leads to poorer image quality at high energies and reduced collection area after event selection. Conversely, in LZA mode, the smaller images from more distant air showers result in better image quality and more accurate direction reconstruction from individual telescopes, compensating for the less effective stereoscopic reconstruction.  }
Figure \ref{fig1:c} shows the relationship between the average {\it MISS} and the impact parameter. Due to the smaller physical size of the images at large zenith angles, the corresponding {\it MISS} is much smaller. Even at an impact parameter of $600~\rm m$, the SDP accuracy obtained from the telescope is around 0.03°, enabling very good direction reconstructions.
Considering the actual geographic conditions, the distance between different cells should be within the range of 360 meters to 410 meters. Therefore, we will use 400 meters as the basis for the following discussion.

\subsection{Offset performance at  LZA} 
\begin{figure}
    \centering
    \includegraphics[width=0.5\textwidth]{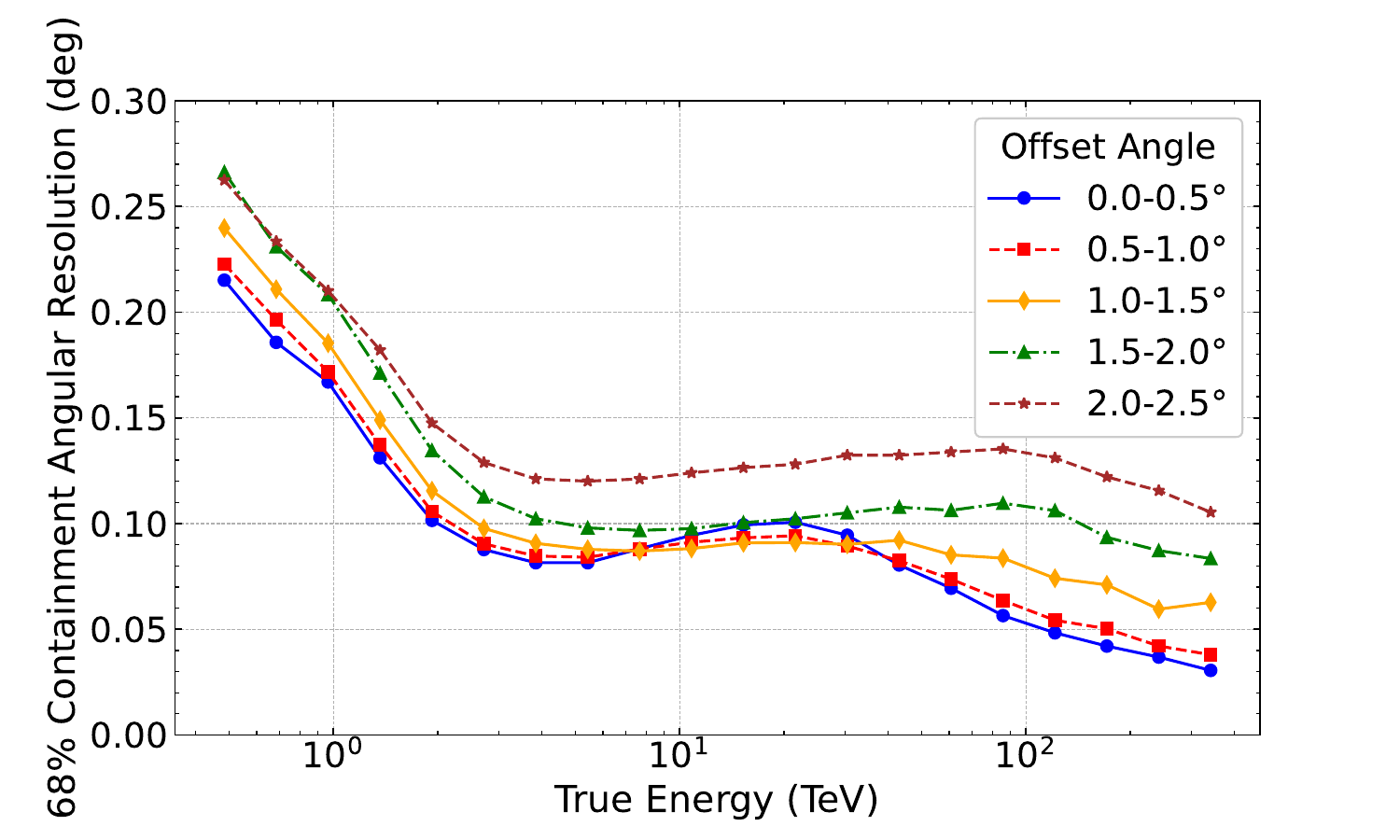}
    \caption{Angular resolution for different offset angles (Zenith Angle $60^{\circ}$).}
    \label{fig:angu_off}
\end{figure}

For LACT telescope, due to the use of the Davies-Cotton single mirror design, its optical performance degrades significantly with offset compared to ASTRI \citep{giro2017first} and CTA-SST, which use a dual mirror design. {Therefore, unlike ASTRI, which maintains a relatively uniform angular resolution within at least 3° off-axis \citep{vercellone2022astri}, LACT's angular resolution significantly deteriorates as the off-axis angle increases.} However, at large zenith angles, the image becomes smaller, which can reduce the degradation in off-axis observation. This makes large zenith angles more suitable for observing extended sources.
Also, the smaller images allow our cameras to have a smaller field of view without compromising performance. 
The off-axis performance was determined using MC diffuse gamma events. In Figure \ref{fig:angu_off},  we show the angular resolution at different offsets for large zenith angles. It can be seen that when the offset is less than 2 degrees, we can still achieve a reasonable angular resolution (better than $0.1^{\circ}$).

\subsection{Overall performance and discussion}

\begin{figure}
    \centering
    \includegraphics[width=0.5\textwidth]{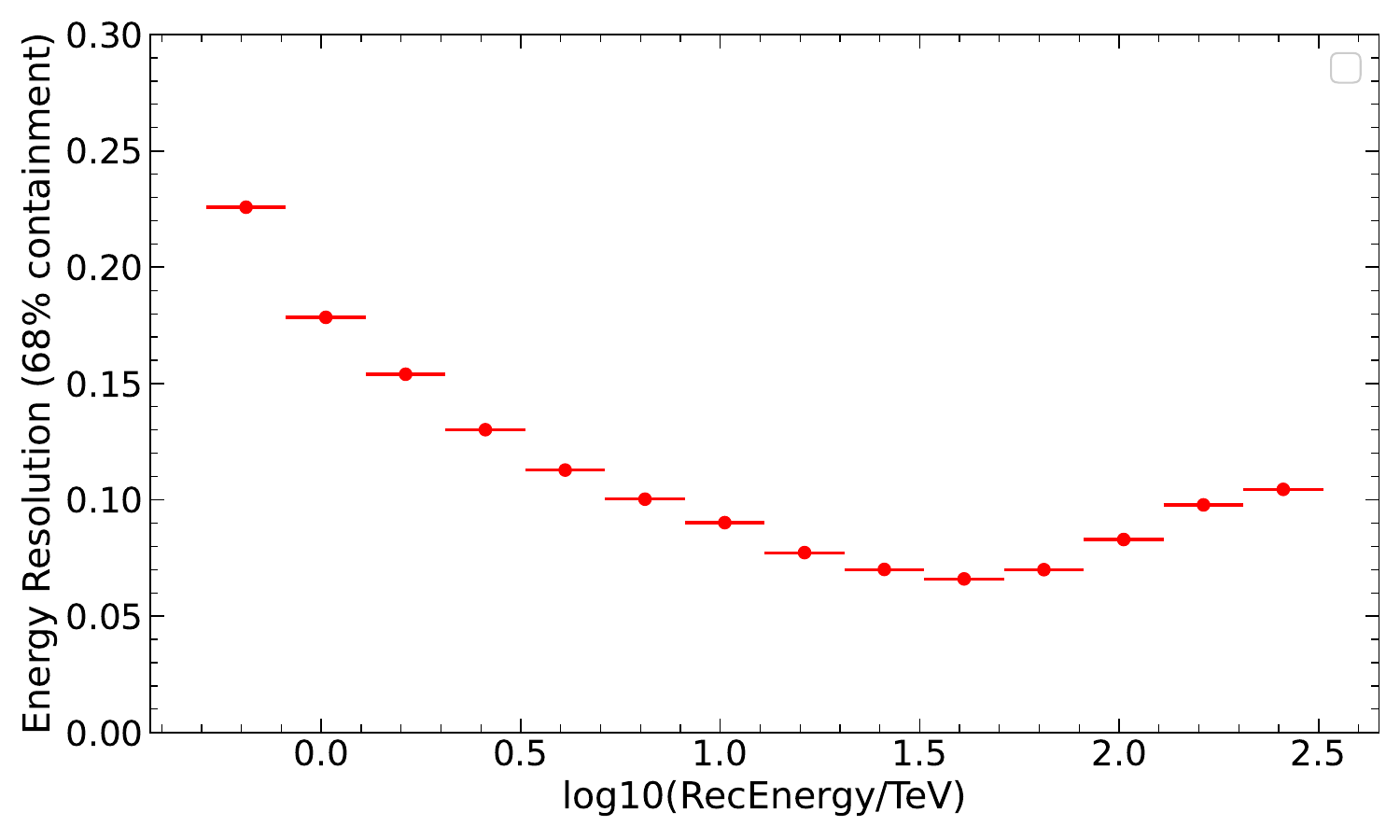}
    \caption{Energy resolution for eight telescopes (Zenith Angle $60 ^{\circ}$)}
    \label{fig:energy_largezenith}
\end{figure}

\begin{figure}
    \centering
    \includegraphics[width=0.5\textwidth]{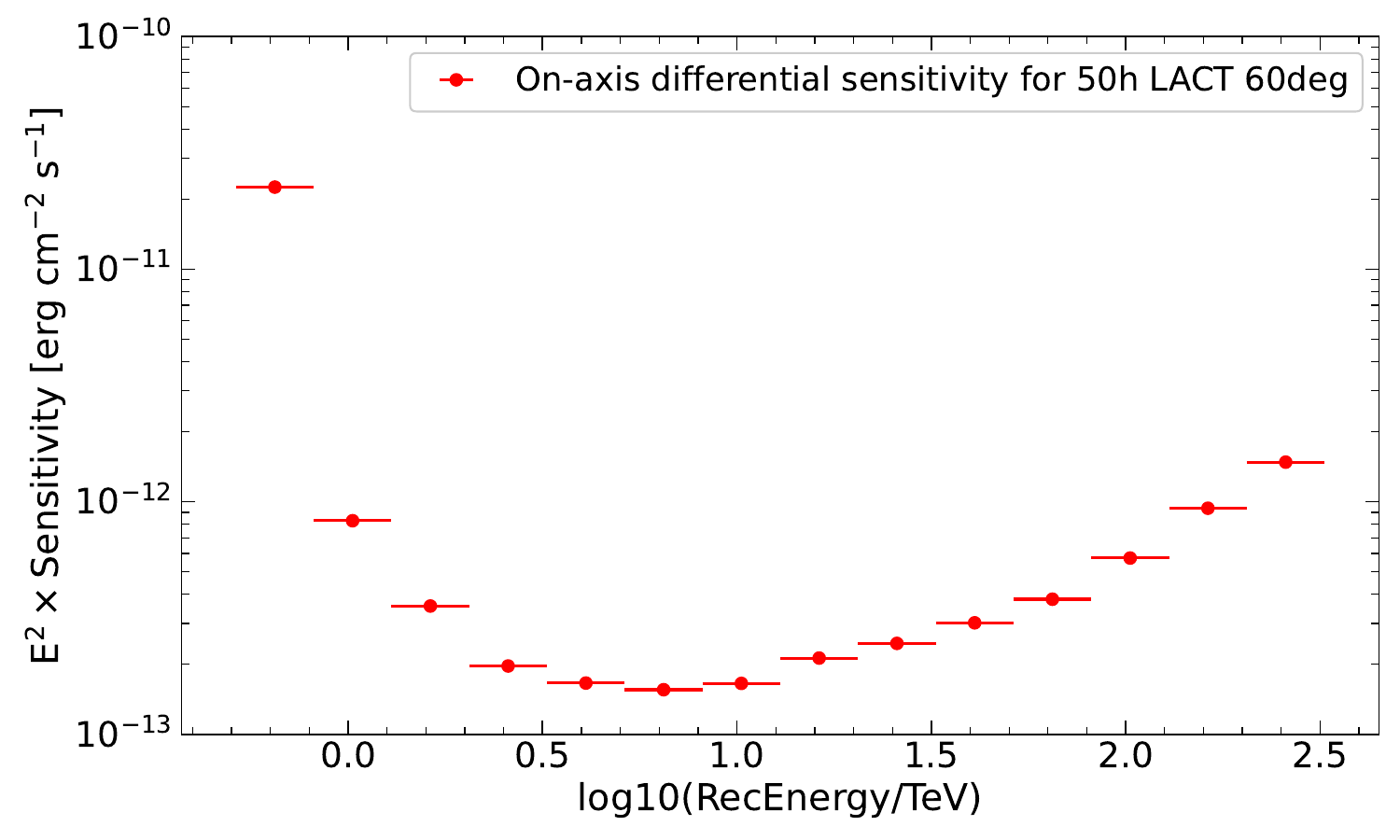}
    \caption{Differential sensitivity for eight telescopes (Zenith Angle $60^{\circ}$)}
    \label{fig:sen_largezenith}
\end{figure}

As mentioned above, energy reconstruction is performed using a RandomForestRegressor model trained on simulated diffuse \gray events, and the result at LZA is shown in Figure \ref{fig:energy_largezenith}. In most energy ranges, the energy resolution is better than $10\%$, enabling accurate spectral measurements. We also calculated the differential sensitivity for 50 hours of observation. The following three conditions are considered: (1) significance greater than 5 (calculated using Eq. 17 from \cite{li1983analysis} and assumed $\rm \alpha = 0.2$). (2) detection of at least 10 photons, and (3) considering the systematic uncertainty of background estimation, we demand \(\rm N_\gamma / N_{\text{background}} > 5\% \). In each energy bin, we optimized the theta cut and particle separation cut to maximize the differential sensitivity. The on-axis differential sensitivity is shown in Figure \ref{fig:sen_largezenith}, owing to the improved angular resolution at large zenith angles and the nearly tripled effective area, we can achieve excellent sensitivity. Compared to the sensitivity of existing IACTs at $20^{\circ}$ zenith angle, LACT's eight-telescope subarray at large zenith angles demonstrates approximately ten times better sensitivity above $30 ~\rm TeV$, approaching that of CTA-South. This exceptional sensitivity allows LACT to achieve significant results in a relatively short observation time.

It's important to note that since the layout of the eight telescopes is not perfectly symmetrical, our performance is likely to depend on the azimuth angle. This dependency might be more pronounced at large zenith angles, necessitating further research.

Moreover, in the large zenith angle observation mode, LHAASO can still provide valuable information. LHAASO-KM2A consists of two main components: the Electromagnetic particle Detectors (ED) and the Muon Detectors (MD). The ED is designed to measure the density and arrival time of electromagnetic particles from extensive air showers \citep{zhao2014design}, while the MD plays a crucial role in discriminating cosmic ray background \citep{wang2020measuring}, thus improving the detection sensitivity for gamma rays. Due to the absorption of electromagnetic particles at LZA, the ED detectors of KM2A face challenges in working effectively. However, the MD detectors remain functional under these conditions.
Considering that LACT can provide excellent angular resolution and core position accuracy for KM2A, KM2A can still play a significant role in particle separation at LZA. Further research is ongoing, but incorporating muon detector information will enhance LACT's particle separation capabilities.

Also, as mentioned in \cite{peresano2018very}, for large zenith angle observation, the accumulation of more clouds and dust makes atmospheric monitoring more crucial. At the LHAASO site, using lasers to monitor the atmosphere up to about 50 km has proven to be highly effective \citep{sun2023properties}. More monitoring equipment will also be installed to assist LACT with calibration.

\section{Small zenith angle performance}
Traditional IACTs typically observe at zenith angles below $50^{\circ}$. Compared to the LZA mode, small zenith angles allow for a lower energy threshold. Therefore, in addition to ensuring the ultra-high energy observations at LZA, LACT also aims to achieve synergy with WCDA, particularly in time-domain \gray astronomy and extragalactic astronomy, such as the detection of GRBs \citep{lhaaso2023very} and Blazars.
In contrast to LZA observations, small zenith angle observations benefit from a relatively close separation between telescopes to achieve optimal performance. Based on the optimization for large zenith angles, we divided the 32 telescopes into 8 well-separated cells, with each cell containing four closely spaced telescopes.  In subsequent analyses, we set the distance between different cells to 400 meters.
\subsection{Optimize the individual CELL}
\begin{figure}
    \centering
    \includegraphics[width=0.5\textwidth]{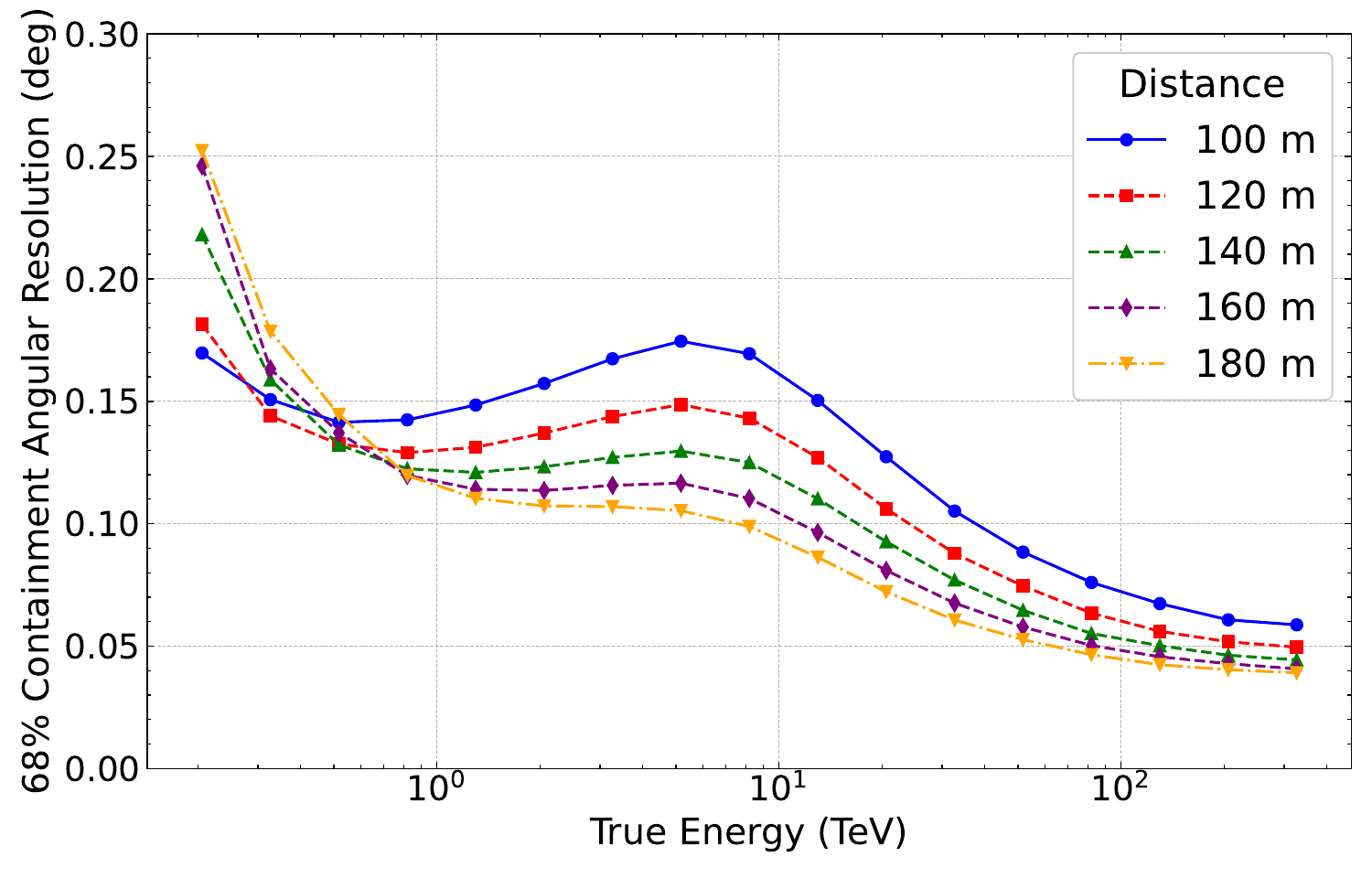}
    \caption{Angular resolution for for CELLs with different side lengths (Zenith Angle $20^{\circ}  $).}
    \label{fig:ang_cell}
\end{figure}

\begin{figure}
    \centering
    \includegraphics[width=0.5\textwidth]{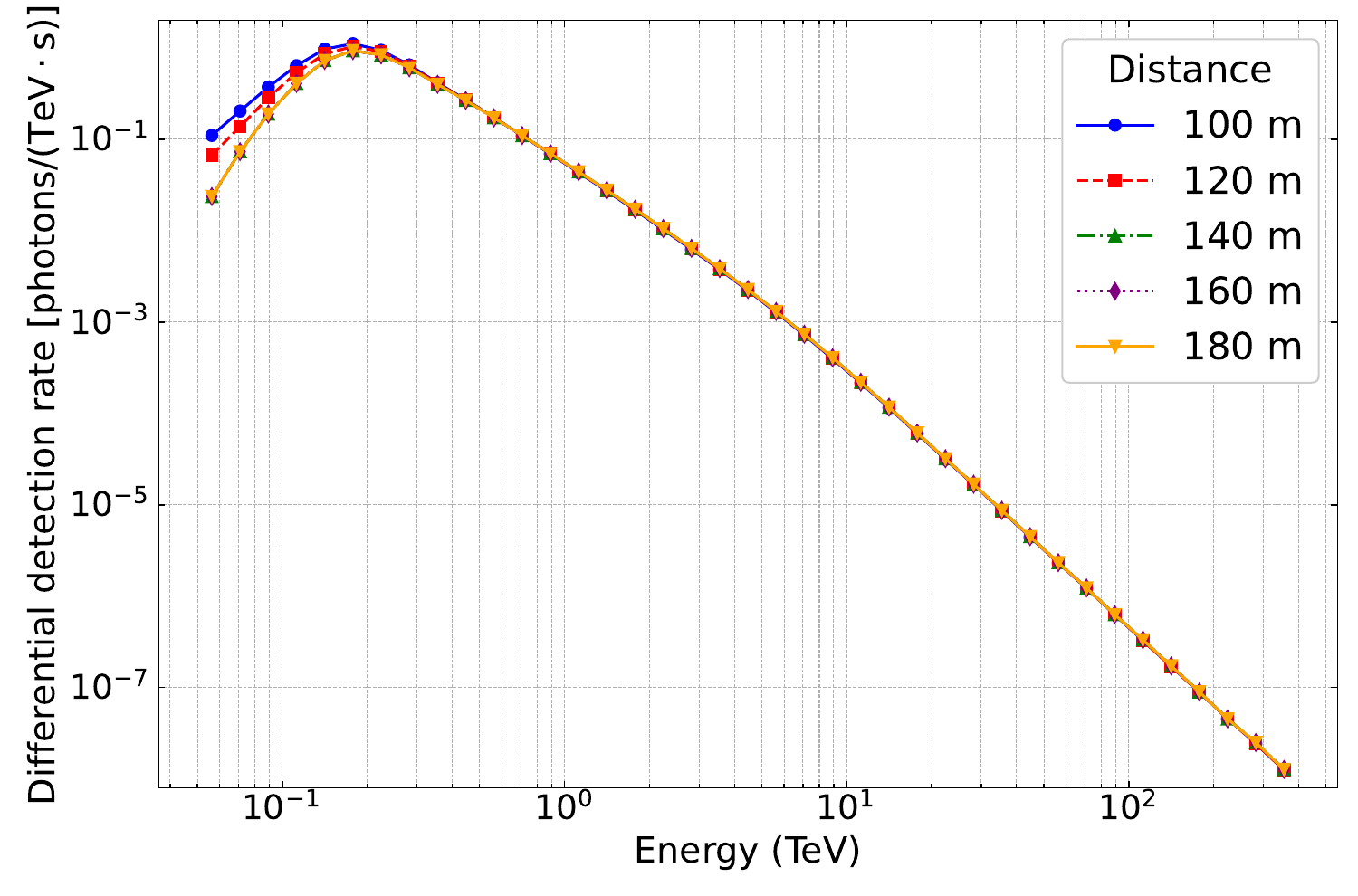}
    \caption{ Differential detection rate for CELLs with different side lengths (Zenith Angle $20^{\circ}$).}
    \label{fig:rate_cell}
\end{figure}

\begin{figure}
    \centering
    \includegraphics[width=0.5\textwidth]{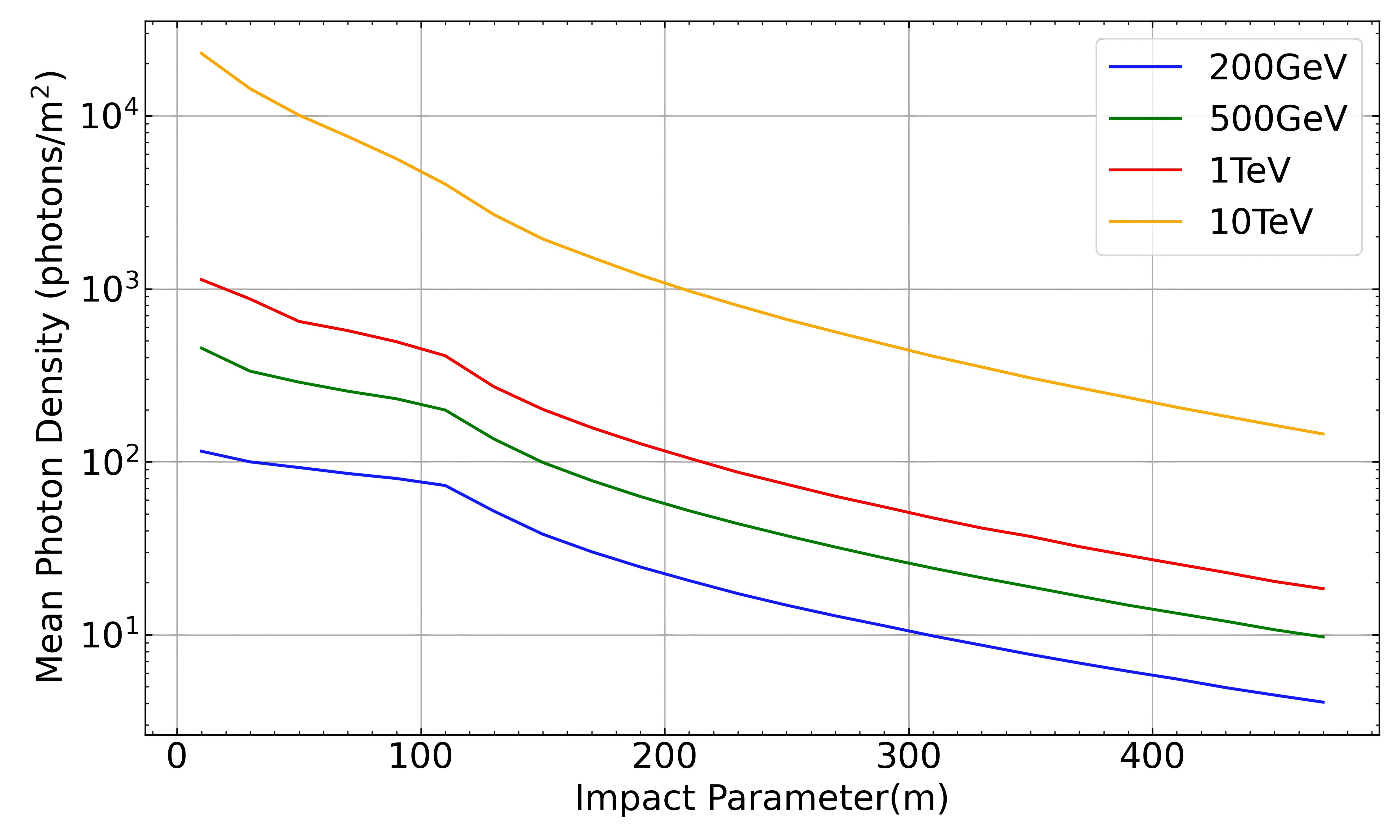}
    \caption{
The lateral distributions of $200~\rm GeV$, $500~\rm GeV$, $1~\rm TeV$, and $10~\rm TeV$ photons obtained from CORSIKA simulations at an altitude of 4400 meters above sea level, without accounting for atmospheric absorption.}
    
\label{fig:lateral_dis}
\end{figure}

Firstly, we consider the performance of a single CELL. Given the large distances between CELLs, the array's performance at low energies will be similar to the arithmetical summation of 8 individual CELLs. A H.E.S.S-like squared CELL is considered, and we investigated the performance with side lengths of $100 ~\rm m$, $120~ \rm m$, $140 ~\rm m$, $160 ~\rm m$, and $180 ~\rm m$, respectively. The angular resolution and differential detection rate are shown in Figure \ref{fig:ang_cell} and Figure \ref{fig:rate_cell}. The differential detection rate \( R(E) \) is computed using the formula \( R(E) = \Phi(E) \times A(E) \), where \( \Phi(E) \) represents the primary \gray differential energy spectrum of the Crab Nebula \citep{aharonian2004crab}, and \( A(E) \) is the collection area. From the differential detection rate, we can determine the energy threshold, which is typically defined as the energy corresponding to the maximum differential detection rate. Interestingly, Figure \ref{fig:rate_cell} reveals that closer distances between telescopes did not result in lower energy thresholds. For CELLs with different side lengths, the corresponding energy thresholds all converged around 200 GeV. To understand this phenomenon, we examined the lateral photon distribution for different energies at LHAASO's altitude of 4410 meters in Figure \ref{fig:lateral_dis}.
 
 At LHAASO's high altitude, {the lateral distribution of photons is steeper compared to lower altitudes. Therefore, for several TeVs \gray showers, there is no so-called light pool \citep{volk2009imaging} present. As the energy decreases, the lateral distribution becomes flatter. Only below $200 ~\rm GeV$ will \gray shower produce a light pool approximately 120 meters in radius.} Within the confines of a light pool, the photon density remains essentially constant. Therefore, closer distances cannot effectively increase the collection area below $200~\rm GeV$, which results in a nearly universal energy threshold. 

\begin{figure}
    \centering
    \includegraphics[width=0.5\textwidth]{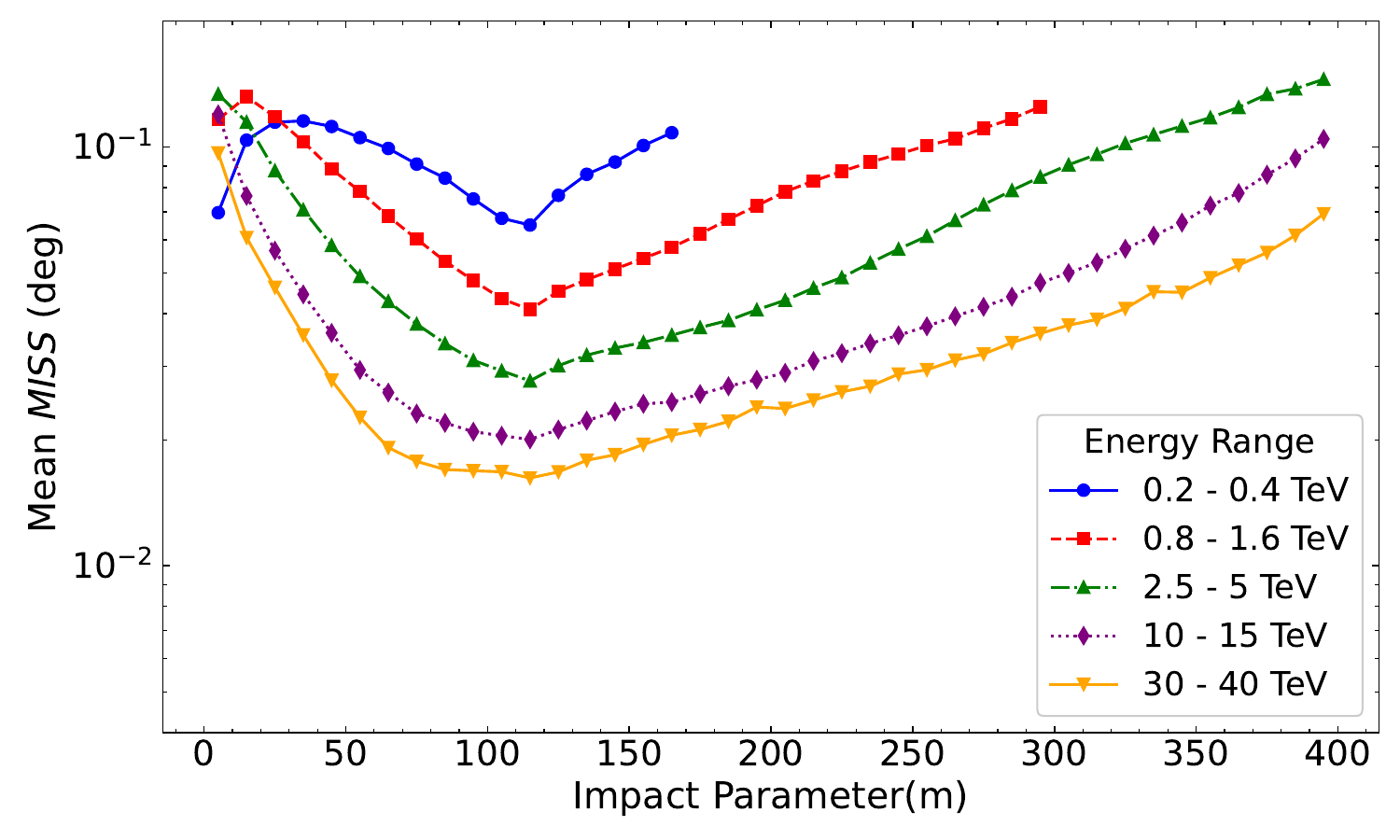}
    \caption{
Mean {\it MISS} versus impact parameter at different energies for a $20^{\circ}$ zenith angle.}
    \label{fig:miss_cell}
\end{figure}

\begin{figure}
    \centering
    \includegraphics[width=0.5\textwidth]{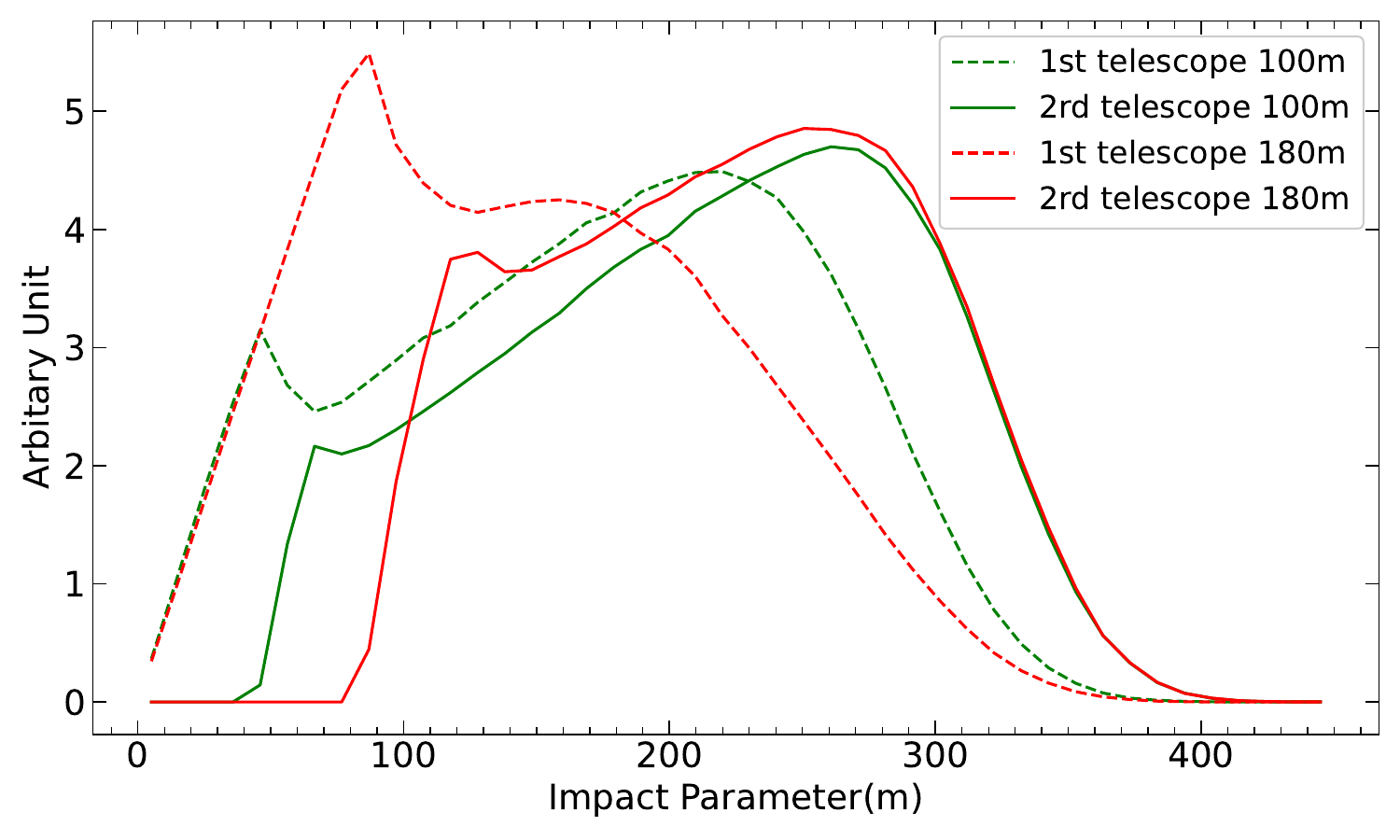}
    \caption{Distribution of impact parameters for the nearest and second nearest telescopes for $2.5 ~\rm TeV$ to $5 ~\rm TeV$ energies, comparing CELLs with side lengths of 180 meters and 100 meters.}
    \label{fig:impact_dis}
\end{figure}

Additionally, for energies above $800 ~\rm GeV$, CELLs with larger side lengths have significantly better angular resolution. Figure \ref{fig:miss_cell} shows the relationship between {\it MISS} and impact parameters at different energies, indicating that the distance corresponding to the minimum {\it MISS} is around 120 meters for most energy ranges. As the impact parameter increases, {\it MISS} initially decreases and then increases. This is because, at smaller impact parameters, geometric factors cause the image in the telescope to become more elongated, resulting in better accuracy. However, as the impact parameter continues to grow, the image moves closer to the edge of the camera, and leakage degrades the imaging quality.
For high-energy events, more events will fall outside the CELL. Therefore, larger side lengths allow more telescopes to be positioned where reconstruction accuracy is higher, resulting in better angular resolution.
This can be verified through Figure \ref{fig:impact_dis}, which shows the distribution of impact parameters for the nearest and second nearest telescopes for $2.5 ~\rm TeV$ to $5 ~\rm TeV$ \gray events, comparing side lengths of 100 meters and 180 meters. We can see that at 180 meters, the nearest telescopes are more frequently positioned between 70 and 200 meters, a range that offers optimal reconstruction accuracy.

\subsection{Comparison of the all telescopes}

\begin{figure}
    \centering
    \includegraphics[width=0.5\textwidth]{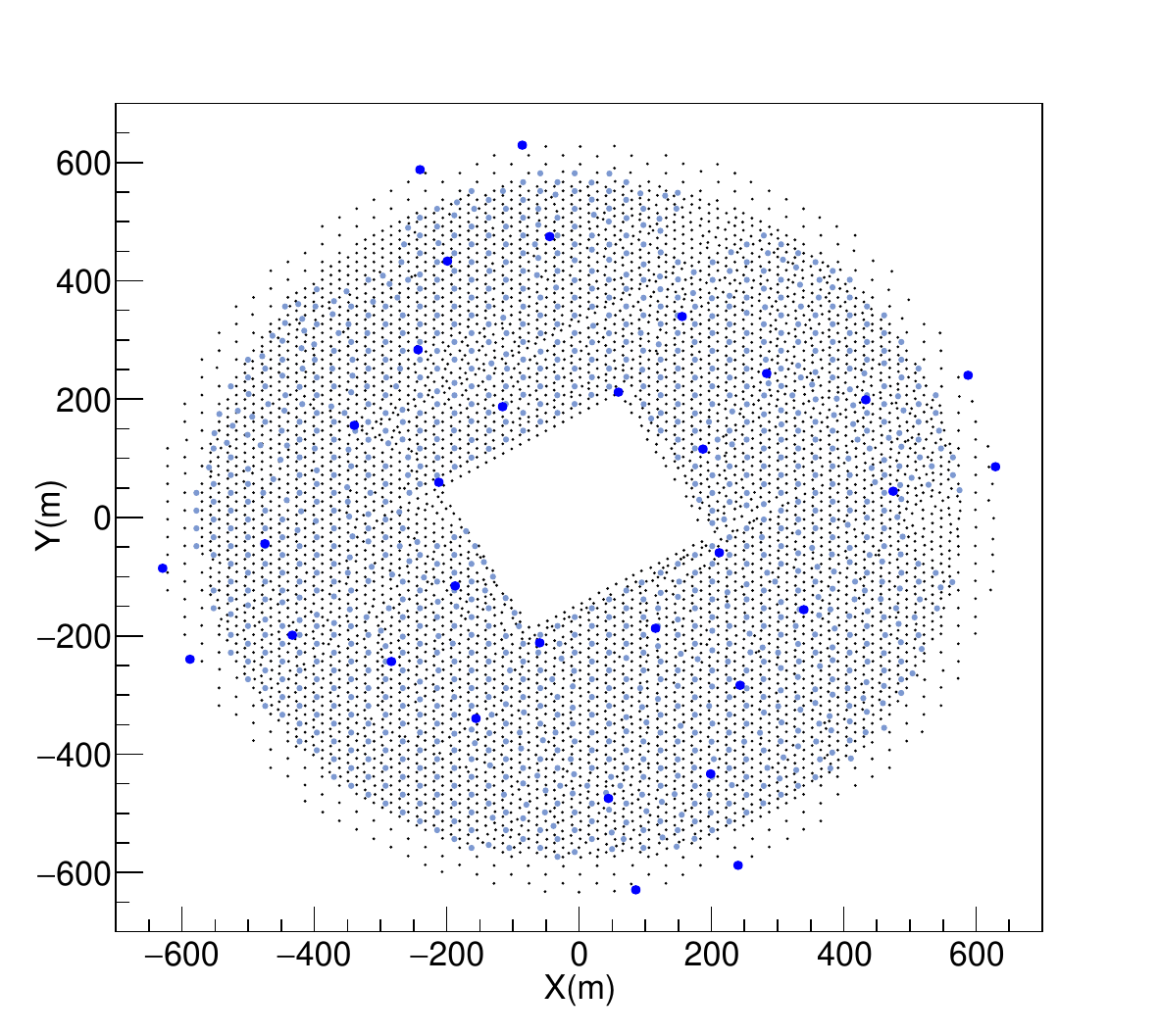}
    \caption{{The layout of LACT and LHAASO. The side length of CELL is set to be 160m. The small black dots represent the electromagnetic particle detectors(EDs) of LHAASO-KM2A, the light blue small circles represent the muon detectors(MDs), and the blue circles represent the LACT telescopes.}}

\label{fig:lact}
\end{figure}

Besides individual CELLs, we also investigate the performance of 32 telescopes with various CELL side lengths. Figure \ref{fig:lact} shows the layout for a side length of 160 meters.

Compared to individual CELLs, the performance differences for various CELL side lengths become much smaller when considering all 32 telescopes.

\begin{figure}
    \centering
    \includegraphics[width=0.5\textwidth]{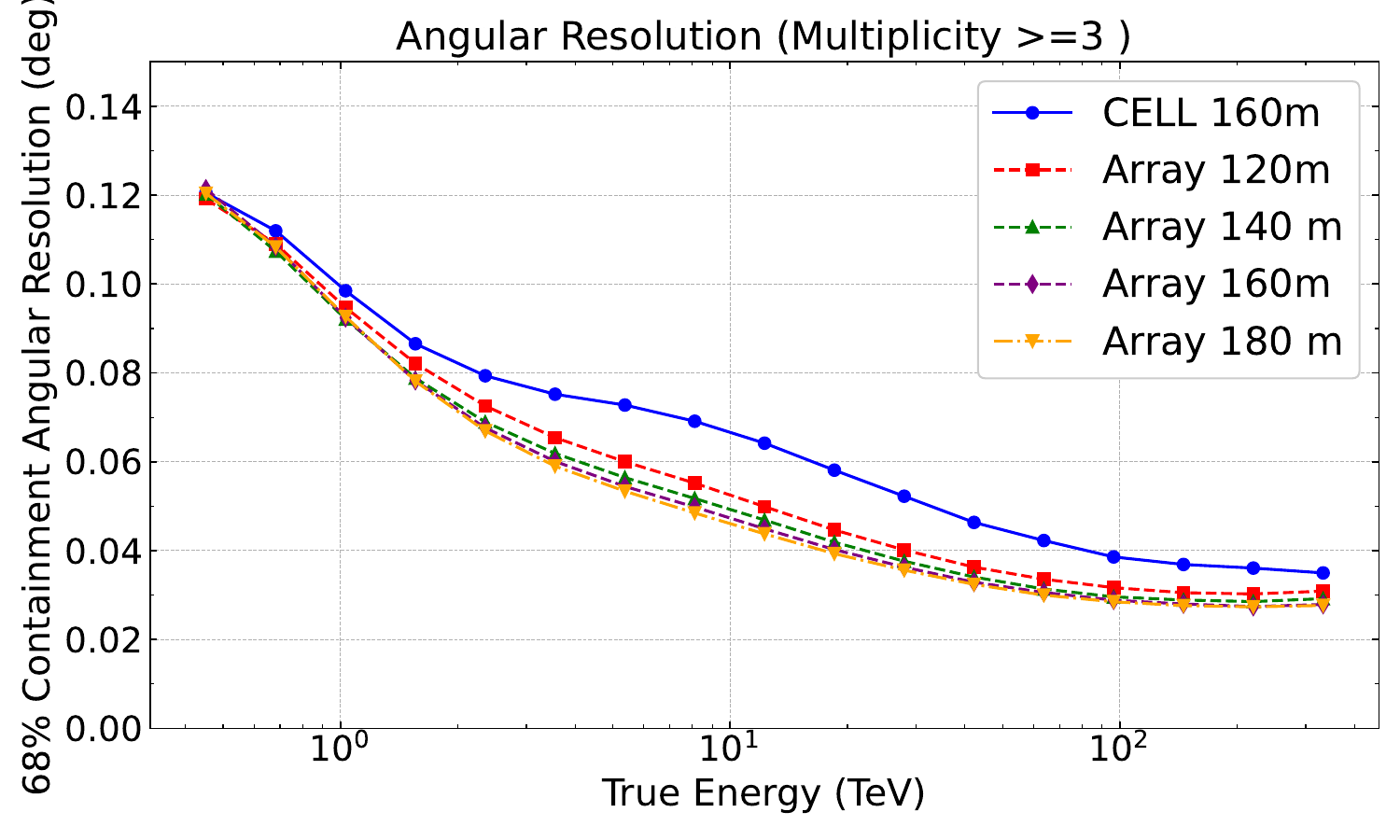}
    \caption{{Angular resolution for different configurations: "CELL" represents four square-like telescopes, while "Array" represents all 32 telescopes. The plot compares the angular resolution for various array side lengths, including the angular resolution for a single CELL with a side length of 160 meters. In order to have a relatively strict cut, we require the multiplicity to be  greater than 3.}}
    \label{fig:cell_versus_array}
\end{figure}
This is because, in the full array mode, when an event moves away from one CELL, it gets closer to adjacent CELLs. Therefore, the performance differences of individual CELLs become less significant. In Figure \ref{fig:cell_versus_array}, we also compared the angular resolution between a single CELL and the full array. It can be seen that for energies below $1 ~\rm TeV$, the angular resolution of the CELL and the array are nearly identical, indicating that low-energy events are mostly contained within a single CELL. As the energy increases, "cross-talk" between different CELLs occurs, leading to a significant improvement in the performance of the full array compared to an individual CELL. 
However, above $100 ~\rm TeV$, the difference between the CELL and the full array decreases. This suggests that, due to leakage, the multiplicity of high-energy events is reduced.

\subsection{Overall performance}

\begin{figure}
    \centering
    \includegraphics[width=0.5\textwidth]{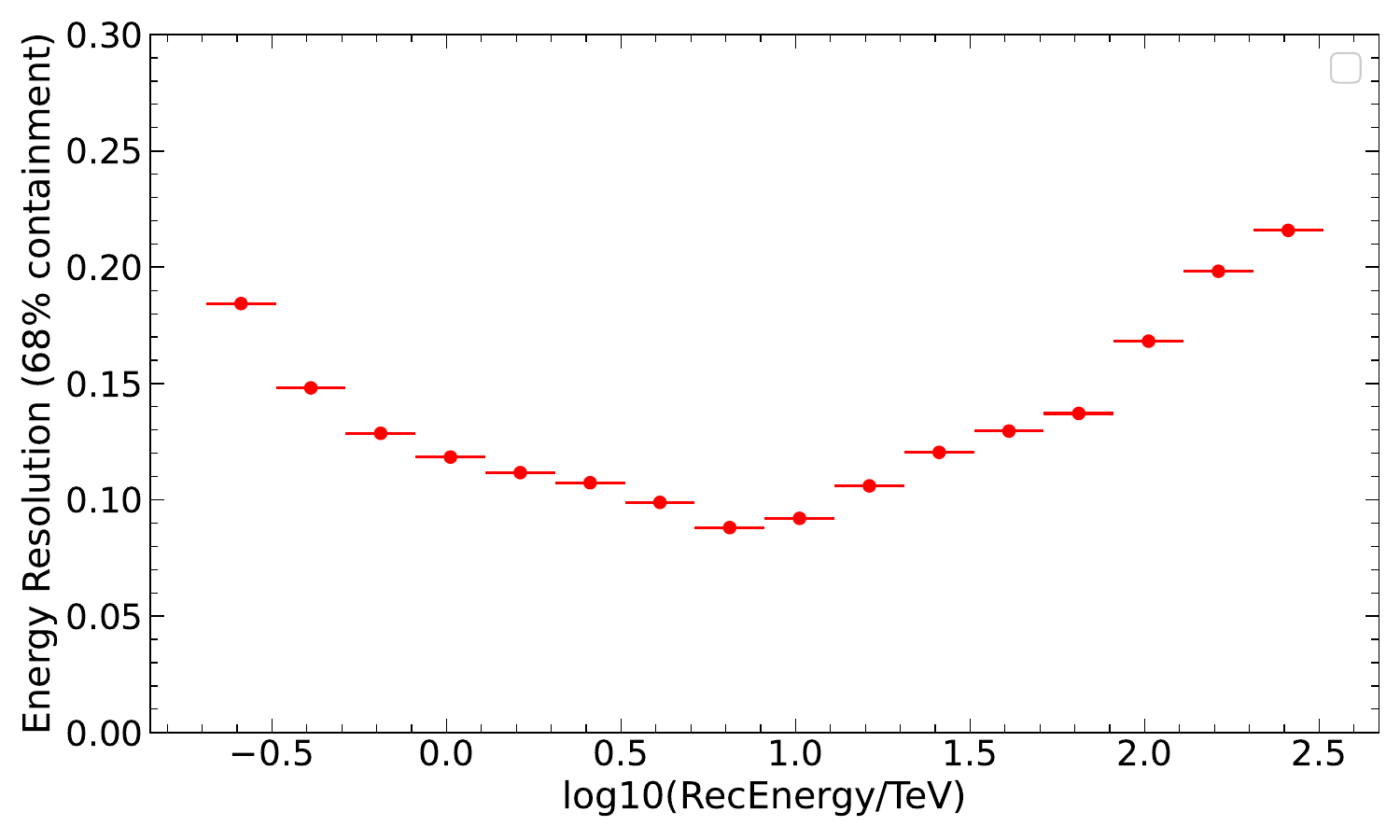}
    \caption{Energy resolution for 32 telescopes. The side length is set to be $140~\rm m$ (Zenith Angle $20 ^{\circ}$). }
    \label{fig:energyres_array}
\end{figure}

\begin{figure}
    \centering
    \includegraphics[width=0.5\textwidth]{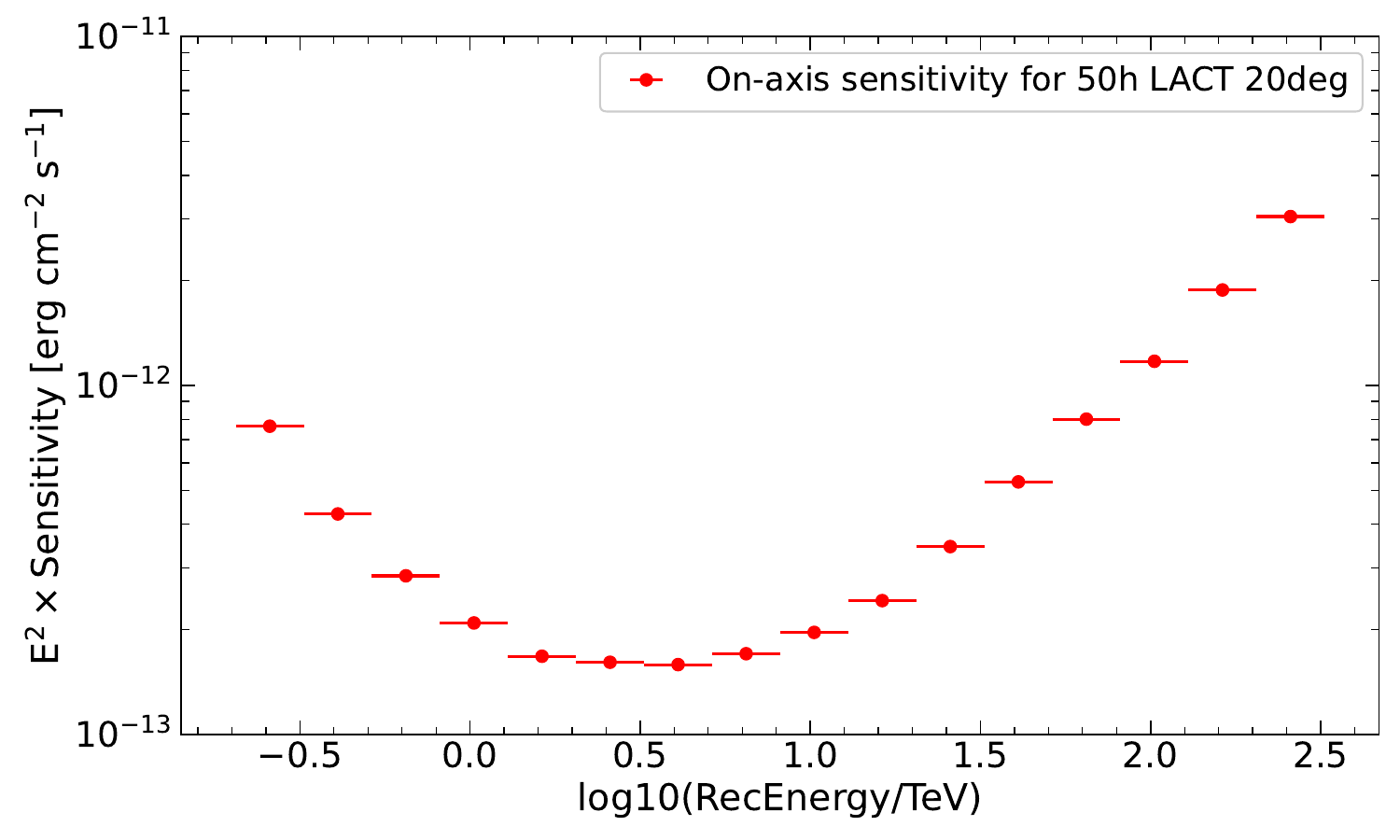}
    \caption{The differential sensitivity of 32 telescopes. The side length is set to be $140~\rm m$(Zenith Angle $20^{\circ}$).}
    \label{fig:sens_array}
\end{figure}

Based on the above analysis, we can conclude that when considering the performance of all 32 telescopes in the LACT array, the performance differences between cells of different side lengths are relatively small. However, for individual cells, larger side lengths offer better performance, particularly in terms of angular resolution for higher energy events. To maintain a balance between overall array performance and the flexibility of individual cell observations, the optimal side length between cells should be between 140m and 180m. Excessively large side lengths would result in the telescopes within different cells being too close to each other, potentially negatively impacting the performance of the entire array. In Figures \ref{fig:energyres_array} and \ref{fig:sens_array}, we present the energy resolution and differential sensitivity of the entire array with a side length of 140m. The energy resolution of the entire array is approximately between 10\% and 20\%, reaching its best value of around 9\% near $5 ~\rm TeV$. Compared to large zenith angles, the energy resolution at high energies is significantly worse due to increased image leakage and less multiplicity. The differential sensitivity of the full LACT array shows remarkable improvement over existing IACTs. At energies of a few hundred GeV, the sensitivity of the full LACT array is approximately twice as good as that of HESS or VERITAS. As the energy increases, this gain becomes even more pronounced. Above a few TeV, LACT is expected to become the most sensitive IACTs in the Northern Hemisphere.

\section{Conclusion}
\begin{figure}
    \centering
    \includegraphics[width=0.5\textwidth]{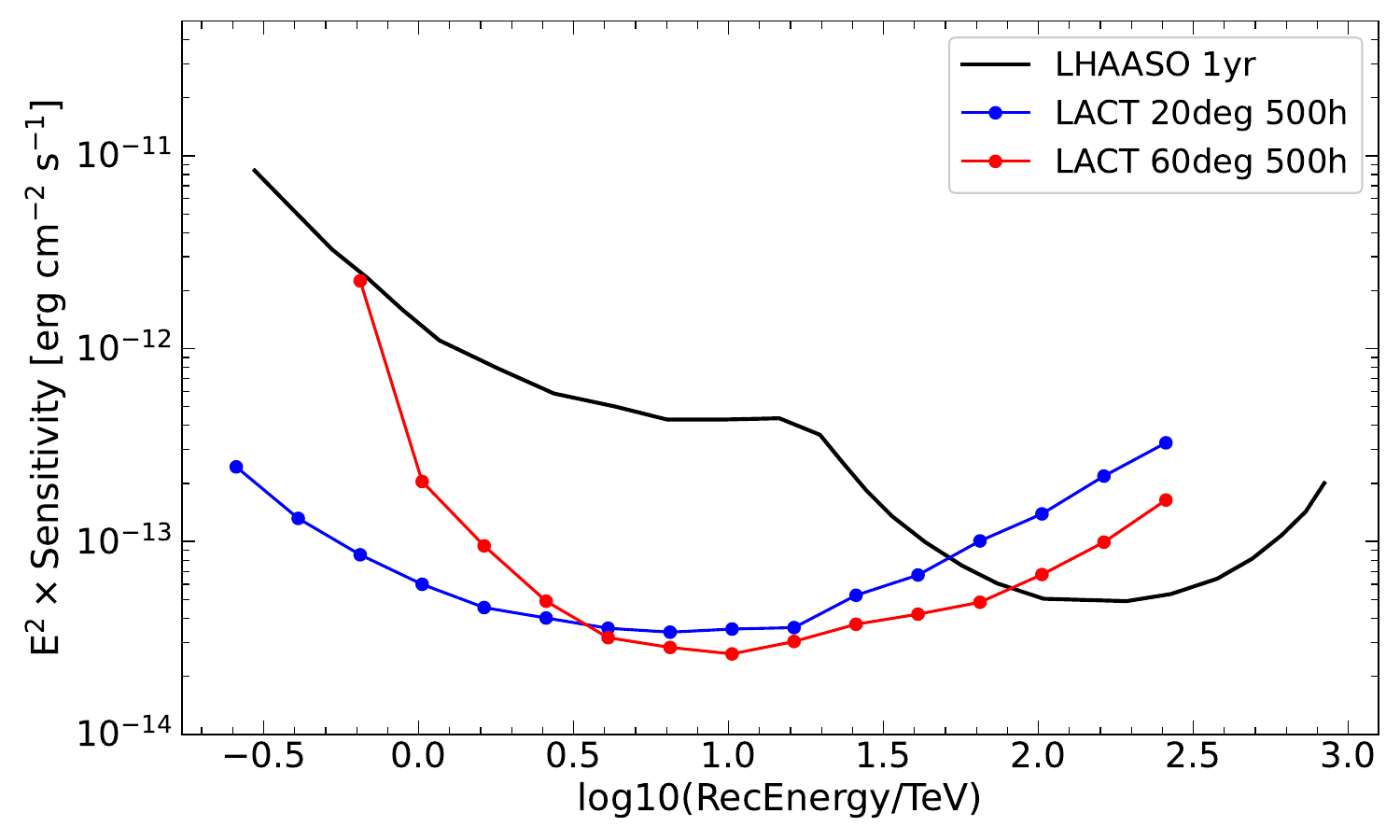}
    \caption{Comparison of the LACT sensitivity under two observation modes after 500 hours of exposure with the LHAASO sensitivity after 1 year. In the 20° observation mode, LACT includes all 32 telescopes in the full array, while in the 60° mode, it consists of a subarray with 8 telescopes.}
    \label{fig:sens_500h}
\end{figure}

The prime scientific objective for LACT is to perform synergy with LHAASO. This objective presents two aspects.  The first is to collaborate with LHAASO-KM2A and conduct deep observations of ultra-high-energy \gray sources and study their morphology in detail, thereby confirming the existence of PeVatrons. To achieve this, we will use the large zenith angle observation mode to increase our effective area. In this paper, we studied the performance of eight telescopes at a 60° zenith angle. Compared to a 20° zenith angle, both the effective area and angular resolution are significantly improved. This eight-telescope subarray can achieve sensitivity close to that of CTA-South and can work synergistically with LHAASO-KM2A. Simultaneously, LACT's four eight-telescope subarrays can observe different sources, significantly increasing the observation time for each source. This capability allows for deep observations of important targets, which could provide detailed spectral and morphological studies of PeVatron candidates. In Figure \ref{fig:sens_500h}, {we present a comparison of the differential sensitivity of LACT after 500 hours in two modes with that of LHAASO after one year. In both modes, LACT demonstrates sensitivity comparable to LHAASO. Notably, in the large zenith angle mode, LACT exhibits better differential sensitivity below 100 TeV.  This positions LACT as a powerful complement to LHAASO, with the potential to significantly enhance our understanding of high-energy astrophysical phenomena
.}
In Table \ref{time},  we have calculated the observation times for several important sources from October 1, 2024, to April 1, 2025, which is considered a good observation period for LACT. The table includes observations at zenith angles below 50° and between 50-70°. From the table, it is evident that for these sources, we can achieve substantial observation times at large zenith angles. Notably, sources such as the Galactic Center can only be observed effectively at large zenith angles, highlighting the importance of LZA observations for expanding LACT's sky coverage and scientific reach.

\begin{table*}
    \centering
    \caption{{Observation times for specific sources by LACT between October 1, 2024, and April 1, 2025, categorized by zenith angles below 50° and between 50-70°. This calculation does not take weather conditions into account and represents an ideal scenario.}}
    \begin{tabular}{lcccccc}
        \toprule
        Source & RA & DEC & 0--50$^\circ$ & 50--70$^\circ$ \\ 
        \midrule
        SS433 & 19h10m37s & +05d02m13s & 75h & 152h \\ 
        J1908+0621 & 19h08m12s & +06d21m0s & 76.25h & 154.75h \\ 
        Galactic center & 17h45m39.6s & -29d0m22s & 0h & 37h \\ 
        J1825-134 & 18h25m49s & -13d46m35s & 2.5h & 99h \\ 
        J2226+6057 & 22h27m0s & +60d57m & 386h & 371h \\ 
        cygnus & 20h31m33s & +41d34m38s & 217h & 233h \\ 
        \bottomrule
    \end{tabular}
    \label{time}
\end{table*}

The second motivation is to achieve a lower energy threshold at small zenith angles and collaborate with LHAASO-WCDA on various scientific topics.  With the construction of next-generation Cherenkov telescopes like CTA and ASTRI, there is a growing desire for the capacity to instantaneously follow up on transient phenomena and continuously monitor them \citep{lee2022performance}. The energy threshold of the entire LACT array is around $200 ~\rm GeV$ at a $20^{\circ}$ zenith angle, allowing for a well-connected observed energy spectrum with Fermi-LAT and excellent synergy with LHAASO-WCDA. This capability is crucial for studying transient events such as gamma-ray bursts (GRBs) and active galactic nuclei (AGNs).

In subsequent studies, we will further investigate the synergy between LHAASO and LACT.
Such synergy is not only in the scientific cases mentioned above but also in experiments themselves such as the joint event reconstructions using different detectors of LHAASO and LACT.  As pointed out in previous research \citep{zhang2024prospects}, during long-term observations of extended sources, the ability to discriminate particles is crucial due to the background count far exceeding that of point sources. The inclusion of KM2A's muon detector can significantly enhance the performance of LACT in long-term observations. Additionally, it's important to note that the simulation parameters currently used do not exactly match those of the actual LACT telescope. By the end of 2024, the first LACT prototype is scheduled to be completed. The observational data from this prototype will enhance our understanding of the simulations, and future Monte Carlo simulations can then be used to validate our analysis methods and derive more realistic performance curves. 

\section*{Acknowledgements}
Rui-zhi Yang is supported by the NSFC under grants 12041305 and the national youth thousand talents program in China. This work is also supported by Sichuan Science and Technology 
Department, Institute of High Energy Physics through the
grants of 2023YFSY0014 and E25156U1, respectively. This study is also supported by the following grants: the Sichuan Province Science Foundation for Distin-guished Young Scholars under grant No. 2022JDJQ0043; the Sichuan Science and Technology Department under grant No. 2023YFSY0014; the Xiejialin Foundation of IHEP under grant No. E2546IU2; the National Natural Science Foundation of China under grants No. 12261141691; the Innovation Project of IHEP under grant No. E25451U2.
\bibliographystyle{dcu}
\bibliography{example}
\end{document}